\documentclass[10pt,aps,prl,twocolumn,notitlepage,floatfix,showpacs,shownokeys,preprintnumbers,nofootinbib,tightenlines,longbibliography]{revtex4-1}
\pdfoutput=1
\usepackage[colorlinks=true,citecolor=blue,linkcolor=blue,breaklinks=true]{hyperref}
\usepackage{amsmath,amssymb,mathtools,tabu,bm}
\usepackage[T1]{fontenc}
\usepackage{textcomp}
\usepackage{lmodern}
\usepackage{enumerate}
\usepackage{cleveref}  
\usepackage{graphicx}   
\usepackage{commath}
\usepackage{slashed} 
\usepackage{comment}            
\usepackage{url}
\usepackage{color}
\usepackage{multirow}
\usepackage{placeins}
\usepackage[dvipsnames]{xcolor}

\newcommand{\myparallel}{{\raisebox{0.5pt}{$\scriptscriptstyle\parallel$}}}
\newcommand\barparenb[1]{\overset{%
   \scalebox{0.2}{${\boldsymbol{(}}\mkern5mu\text{\rule[.5ex]{1.5em}{0.8pt}}\mkern5mu{\boldsymbol{)}}$}}{#1}}

\DeclareFontFamily{U}{mathx}{\hyphenchar\font45}
\DeclareFontShape{U}{mathx}{m}{n}{<-> mathx10}{}
\DeclareSymbolFont{mathx}{U}{mathx}{m}{n}
\DeclareMathAccent{\widebar}{0}{mathx}{"73}

\allowdisplaybreaks

\bibpunct{[}{]}{,}{n}{}{,}

\newcommand{\bd}[1]{{\color{black}#1}}

\usepackage{tikz,xcolor,hyperref}

\definecolor{lime}{HTML}{A6CE39}
\DeclareRobustCommand{\orcidicon}{\hspace{-1mm}
	\begin{tikzpicture}
	\draw[lime, fill=lime] (0,0) 
	circle [radius=0.16] 
	node[white] {{\fontfamily{qag}\selectfont \tiny \,ID}};
	\draw[white, fill=white] (-0.0525,0.095) 
	circle [radius=0.007];
	\end{tikzpicture}
	\hspace{-3mm}
}

\foreach \x in {A, ..., Z}{\expandafter\xdef\csname orcid\x\endcsname{\noexpand\href{https://orcid.org/\csname orcidauthor\x\endcsname}
			{\noexpand\orcidicon}}
}

\begin{document}

\title{Fast Flavor Depolarization of Supernova Neutrinos}
       
\author{Soumya~Bhattacharyya\orcidA{}}
\email{soumya.bhattacharyya@theory.tifr.res.in}
\affiliation{Tata Institute of Fundamental Research,
	Homi Bhabha Road, Mumbai, 400005, India}
	
\author{Basudeb~Dasgupta\orcidB{}}
\email{bdasgupta@theory.tifr.res.in}
\affiliation{Tata Institute of Fundamental Research,
             Homi Bhabha Road, Mumbai, 400005, India}	

\preprint{TIFR/TH/20-33}
\date{September 12, 2020}

\begin{abstract}
Flavor-dependent neutrino emission is critical to the evolution of a supernova and its neutrino signal. In the dense anisotropic interior of the star, neutrino-neutrino forward-scattering can lead to fast collective neutrino oscillations, which has striking consequences. We present a theory of fast flavor depolarization, explaining how neutrino flavor differences become smaller, i.e., depolarize, due to diffusion to smaller angular scales. We show that transverse relaxation determines the epoch of this irreversible depolarization. We give a method to compute the depolarized fluxes, presenting an explicit formula for simple initial conditions, which can be a crucial input for supernova theory and neutrino phenomenology.     
\end{abstract}

\maketitle 

Metronomes sway in lockstep, crickets chirp in a chorus, and neurons fire in sync -- all examples of coordinated action by seemingly unregulated agents~\cite{strogatz2012sync}. Neutrinos emitted by collapsing stars can also exhibit such collective behavior in their quantum mechanical flavor oscillations~\mbox{\cite{Pantaleone:1992xh,Kostelecky:1994dt,Pastor:2001iu,Friedland:2003dv,Bell:2003mg, Duan:2006an,Hannestad:2006nj,Raffelt:2007yz,Raffelt:2007cb,EstebanPretel:2007ec,Dasgupta:2007ws,Dasgupta:2008cu,Dasgupta:2009mg,Friedland:2010sc,Pehlivan:2011hp,Chakraborty:2014nma,Mangano:2014zda,Duan:2014gfa,Dasgupta:2015iia,Birol:2018qhx,Hansen:2019iop,Cervia:2019res}}. Astonishingly, this dense gas of neutrinos can change its flavor at a rate proportional to the neutrino density~\cite{Sawyer:2005jk, Sawyer:2008zs, Sawyer:2015dsa,Chakraborty:2016lct, Dasgupta:2016dbv,Izaguirre:2016gsx,Capozzi:2017gqd,Dasgupta:2017oko, Morinaga:2018aug,Dasgupta:2018ulw,Airen:2018nvp,Abbar:2018beu,Capozzi:2019lso,Martin:2019gxb,Johns:2019izj,Bhattacharyya:2020dhu}, much faster than any individual neutrino. It's as if a marching band outruns Usain Bolt. Such fast evolution may erase the differences between neutrino fluxes, i.e., depolarize in flavor, within picoseconds and over distances smaller than a pinhead. In this \emph{Letter}, we propose a theory of ``fast flavor depolarization'', which has major consequences for supernova (SN) explosions and their signals at neutrino telescopes.

Fast oscillations are a peculiar avatar of neutrino oscillation. \bd{They involve pairwise $\nu_{e}\bar{\nu}_{e}\leftrightarrow \nu_{\mu,\tau}\bar{\nu}_{\mu,\tau}$ conversions~\cite{Sawyer:2005jk, Sawyer:2008zs, Sawyer:2015dsa,Chakraborty:2016lct, Dasgupta:2016dbv,Izaguirre:2016gsx} that proceed} 
at a rate $\sqrt{2}G_{F}n_{\nu}$\,$\sim$\,10\,cm$^{-1}$, proportional to the local neutrino density\,{$\sim$\,$(10^{35}$\,-\,$10^{30})$\,$\text{cm}^{-3}$} at radii {$r\sim(10$\,-\,$100)$\,km} in a SN~\cite{Tamborra:2017ubu}. This rate greatly exceeds the oscillation rate in vacuum $\omega={|\Delta m^{2}|/(2{E})}$\,$\sim$\,km$^{-1}$. (We use $\hbar=c=1$, expressing everything in units of length or time.) As such, fast oscillation is quite insensitive to the size or sign of the neutrino-mass-square difference $\Delta m^{2}$, and stems from an \emph{instability} that can be triggered by any nonzero $\omega$~\cite{Chakraborty:2016lct}.

\begin{figure}[!t]
\includegraphics[width=0.95\columnwidth]{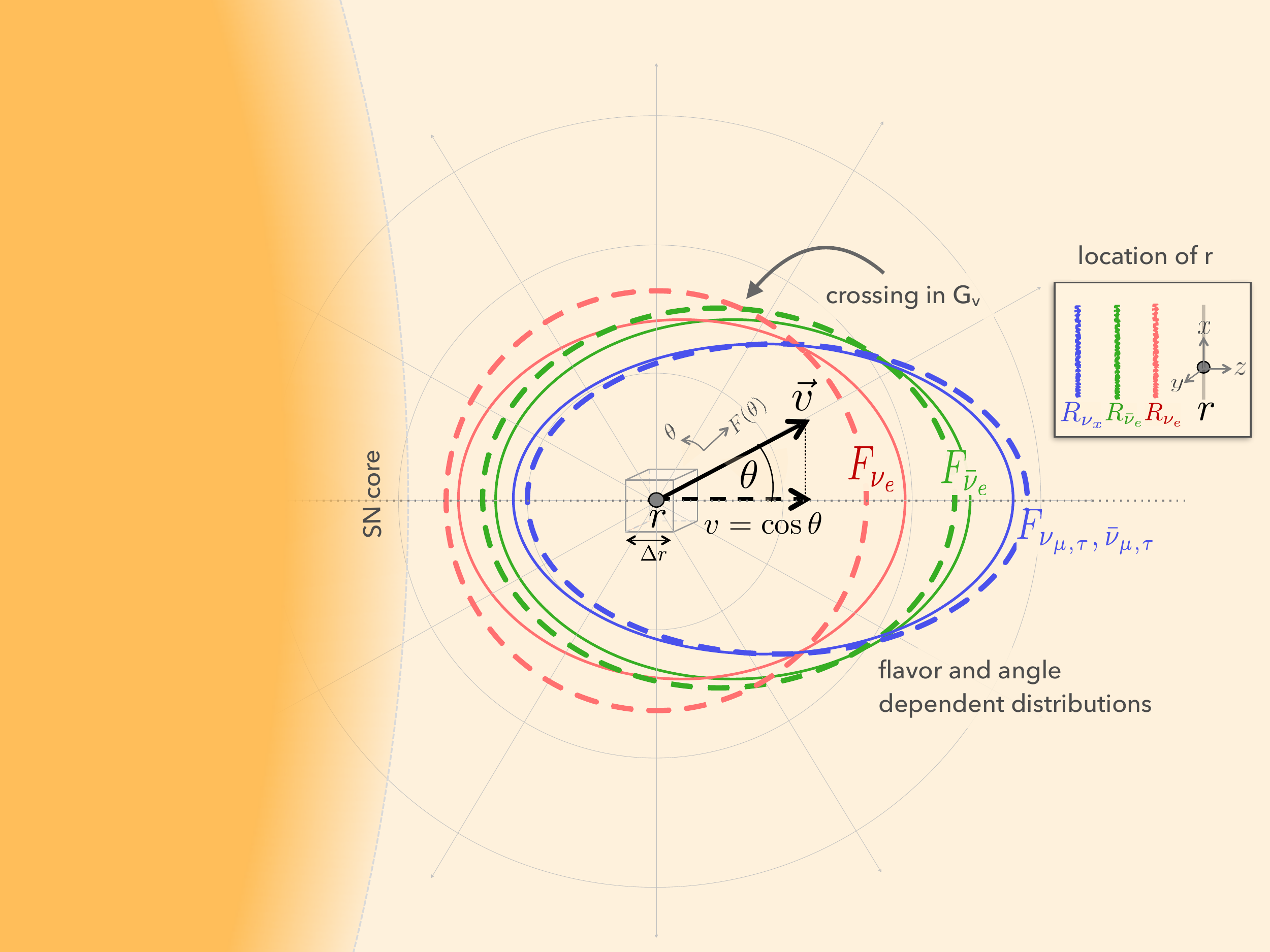}
\caption{{\sc Schematic:}  \bd{SN neutrino decoupling, just above $R_{\alpha}$,} with illustrative polar plots of angle-dependent neutrino distributions $F_{\alpha}$, initially (thick dashed ellipses) with a forward-excess of $\bar\nu_{e}$ (green) over $\nu_{e}$ (red), producing a \emph{crossing}, and of ${\nu}_{x}$ (blue) over $\bar\nu_{e}$, and finally (thin ellipses) their differences reduced due to depolarization.}
\label{fig1}
\end{figure}

Neutrino distributions, $F_{\alpha}[\vec{p}\,]=d^{3}n_{\alpha}/d^{3}{\vec p}$, \bd{vary with direction in a flavor-dependent manner.} Here $\alpha=\barparenb{\nu}_{e,\,\mu,\,\tau}$. If the ${\nu}_{\mu,\tau}$ and $\bar{\nu}_{\mu,\tau}$ flavors are almost identical (hereafter denoted as $\nu_{x}$), as motivated by the much lower $\mu^{\pm}$ and $\tau^{\pm}$ densities than those of $e^{\pm}$, the criterion for instability is met if the $\nu_{e}$ and $\bar\nu_{e}$ distributions are equal along some direction(s)~\cite{Chakraborty:2016lct, Dasgupta:2016dbv,Izaguirre:2016gsx,Capozzi:2017gqd,Dasgupta:2017oko, Morinaga:2018aug,Dasgupta:2018ulw,Airen:2018nvp,Abbar:2018beu,Capozzi:2019lso,Martin:2019gxb,Johns:2019izj,Bhattacharyya:2020dhu}. Fig.\,\ref{fig1} shows a sketch of the decoupling region in the SN. The different neutrino flavors have hierarchical interaction rates, and they kinetically decouple at $R_{\nu_{e}}>R_{\bar\nu_{e}}>R_{{\nu}_{x}}$. In the decoupling region, this can produce relative forward excesses in the fluxes of ${\nu}_{x}$ over $\bar{\nu}_{e}$, and $\bar{\nu}_{e}$ over $\nu_{e}$~\mbox{\cite{Capozzi:2018clo,Shalgar:2019kzy,Morinaga:2019wsv,Abbar:2019zoq,Glas:2019ijo}}, as shown in the schematic polar plots. This allows the $\nu_{e}$ and $\bar\nu_{e}$ distributions to develop a \emph{crossing}, as believed to be required for the fast instability.

Although the triggering and initial growth of fast oscillations are well understood~\cite{Sawyer:2005jk, Sawyer:2008zs, Sawyer:2015dsa,Chakraborty:2016lct, Dasgupta:2016dbv,Izaguirre:2016gsx,Capozzi:2017gqd,Dasgupta:2017oko, Morinaga:2018aug,Dasgupta:2018ulw,Airen:2018nvp,Abbar:2018beu,Capozzi:2019lso,Martin:2019gxb,Johns:2019izj,Bhattacharyya:2020dhu}, owing to complicated nonlinear evolution~\cite{Dasgupta:2017oko,Abbar:2018beu,Bhattacharyya:2020dhu} the final impact is not yet known. Stellar explosion and the neutrino signal are sensitive to the nonlinearly processed flavor-dependent fluxes, and the required neutrino theory prediction of these fluxes is lacking. In this work, we address this crucial theoretical and phenomenological obstacle and pave a clear path forward. We present a theory that explains \emph{how}, \emph{when}, and \emph{to what extent} do the flavor differences change due to fast oscillations.

\begin{figure}[!t]
 	\begin{centering}
 		\includegraphics[width= 0.76\columnwidth]{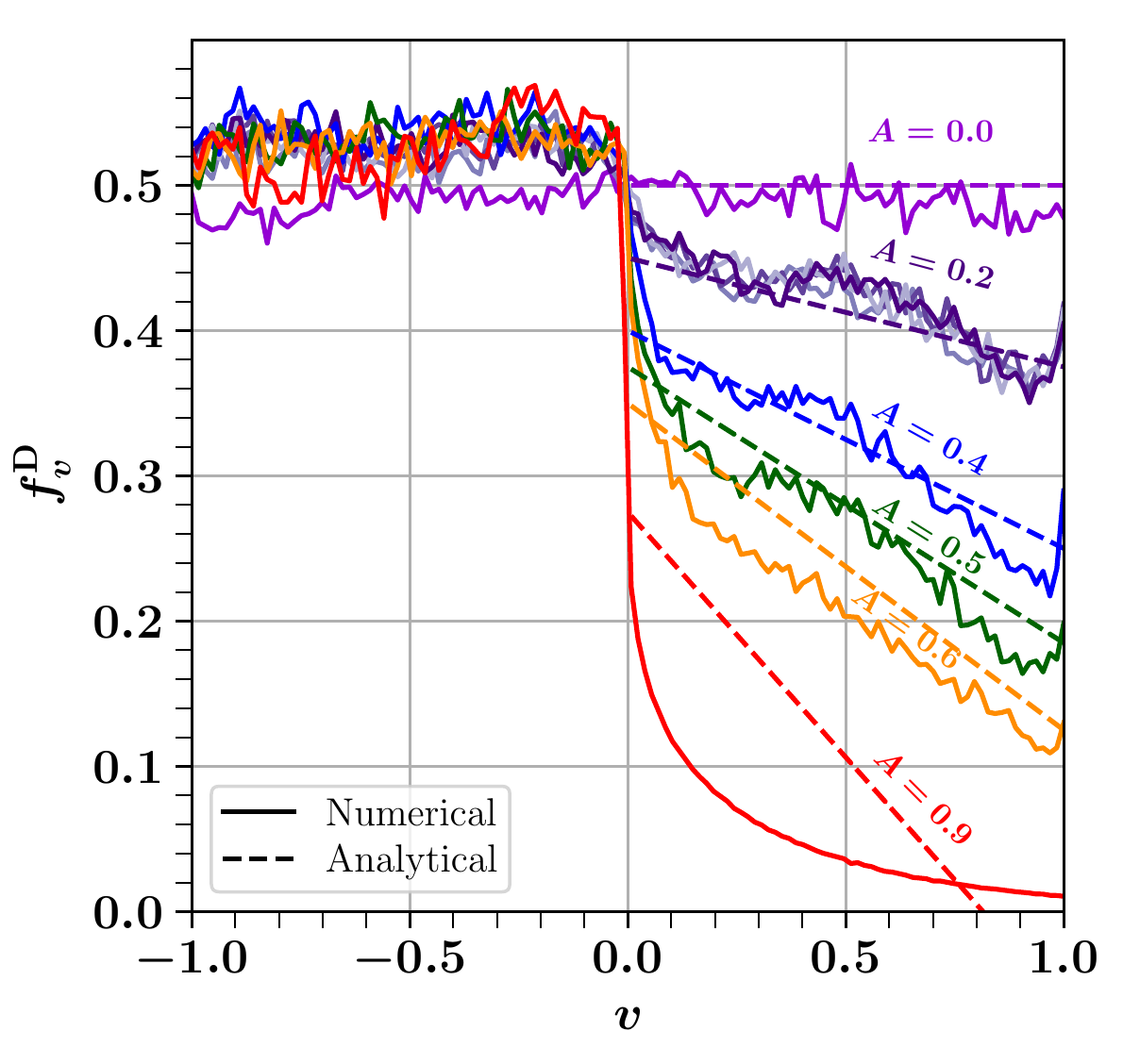}
 		\caption{{\sc Depolarization Factor:} Analytical (dashed) and numerical (solid) results for \bd{coarse-grained} $f^\textrm{D}_v$, as a function of the \bd{radial velocity}, $v=\cos\theta$, for different initial neutrino \bd{ELNs} labeled by their lepton asymmetry $A$. \bd{For $A=0.2$, the different purple lines are for different initial seeds.}}
		\label{fig2}
 	\end{centering}	
 \end{figure}

For two flavors, say $e$ and $\mu$, the final distributions after depolarization can be written as
\begin{align}\label{eq:flux}
F^\textrm{fin}_{\barparenb{\nu}_{e},\,\barparenb{\nu}_{\mu}}[\vec{p}\,]=(1-f^\textrm{D}_{\vec{p}})F^\textrm{ini}_{\barparenb{\nu}_{e},\,\barparenb{\nu}_{\mu}}[\vec{p}\,]+f^\textrm{D}_{\vec{p}}F^\textrm{ini}_{\barparenb{\nu}_{\mu},\,\barparenb{\nu}_{e}}[\vec{p}\,]\,,
\end{align} 
where the depolarization factor $f^\textrm{D}_{\vec{p}}$, which is the same for $\nu$ and $\bar\nu$, is equal to \textonehalf~for perfect equality of distributions and 0 for no change. Values between \textonehalf~and 1 indicate effective flavor conversion. We will present an explicit formula for $f^\textrm{D}_{v}$ [in Eq.\eqref{eq:fd}], assuming an azimuth-symmetric $F$. This result for $f^\textrm{D}_{v}$ is previewed in Fig.\,\ref{fig2}. As predicted analytically, the extent of depolarization depends on the radial velocity $v=\cos\theta$ and lepton asymmetry $A\propto (n_{\nu_{e}}-n_{\bar\nu_{e}})$. In the following, we set up the problem, present our theory that leads to this result, and conclude by discussing the relevance of our results to SN physics and neutrino phenomenology.

\emph{Set-up {\textit{\&}} Notation.---} 
As shown in Fig.\,\ref{fig1}, we consider a small region \bd{of size $\Delta r$} around $r$, just outside radii $R_{\alpha}$ in a SN where ${\cal O}(G_{F}^{2})$ momentum-changing collisions have ceased. In a realistic SN, $R_{\alpha}$\,$\sim$\,km and $(r-R_{\alpha})\ll R_{\alpha}$. The equation for a two-flavor $\lvert\nu\rangle$ with energy-momentum~$(E,\vec{p})$, in a spacetime volume where all macroscopic parameters such as density $n$ are constant, is~\cite{Chakraborty:2016lct, Dasgupta:2016dbv,Capozzi:2019lso,Bhattacharyya:2020dhu}
\begin{align}\label{eom1}
\left(\partial_{t}+\vec{v}.\vec{\nabla}\right)\mathsf{S}_{\omega, \vec{v}} = \left(\mathsf{H}^{\rm vac}_{\omega}+\mathsf{H}^{\rm mat}+\mathsf{H}^{\rm self}_{{\vec v}}\right)\times \mathsf{S}_{\omega, \vec{v}}\,.
\end{align}
Antineutrinos are represented with $\omega=-|\Delta m^{2}|/(2{E})$, extending $\omega$ to negative values. Sans-serif letters denote vectors in flavor space, whose magnitudes are shown in the usual font. E.g., $\mathsf{S}_{\omega, \vec{v}}[\vec{r},t]$, with $\lvert \mathsf{S}_{\omega, \vec{v}}\rvert\equiv S_{\omega, \vec{v}} =1$, is the normalized Bloch vector corresponding to the density matrix $\lvert\nu_{\omega, \vec{v}}\rangle\langle\nu_{\omega, \vec{v}}\rvert$ varying in $(\vec{r},t)$. We work in the flavor basis $\{\hat{\mathsf{e}}_{1},\,\hat{\mathsf{e}}_{2},\,\hat{\mathsf{e}}_{3}\}$, where the longitudinal component along $\hat{\mathsf{e}}_{3}$ is denoted by $(\cdot)^{\myparallel}$ and the transverse by $(\cdot)^{\perp}$. Thus, ${\sf S}^{\myparallel}$ encodes the flavor composition $\lvert\langle\nu_{e}\vert\nu\rangle\rvert^{2}-\lvert\langle\nu_{\mu}\vert\nu\rangle\rvert^{2}$. Note that ${\sf S}^{\myparallel}$ can be negative, but not ${S}^{\myparallel}=\lvert{\sf S}^{\myparallel}\rvert$.
The vector $\mathsf{H}^{\rm vac}_{\omega} = \omega \big(\sin{2\vartheta}, 0, \cos{2\vartheta}\big)$ causes oscillations in vacuum, $\mathsf{H}^{\rm mat}=\sqrt{2} G_{F} (n_{e^{\text{--}}}-n_{e^+})\big(0, 0,1\big)$ gives matter effects, and {$\mathsf{H}^{\rm self}_{{\vec v}} =\int d^3\vec{p}\,'_{\omega',\vec{v}{'}}/(2\pi)^{3}\,g_{\omega', \vec{v}'}\big(1-\vec{v}\cdot\vec{v}{'}\big) \mathsf{S}_{\omega{'}, \vec{v}{'}}$}, with \mbox{$g_{\omega, \vec{v}}=(F_{{\nu}_{e}}-F_{{\nu}_{\mu}})$} for $\omega>0$ and $(F_{\bar{\nu}_{\mu}}-F_{\bar{\nu}_{e}})$ for $\omega<0$, causes collective effects.

In the fast oscillation limit, we neglect the \bd{${\sf H}^{\rm vac}_{\omega}$ and ${\sf H}^{\rm mat}$} in Eq.\eqref{eom1}, compared to \bd{${\sf H}^{\rm self}_{\vec v}$}. The self-term then enters the Hamiltonian only through the  difference  of distributions  integrated  over  \bd{$\omega$}~\cite{Chakraborty:2016lct}, defined by the electron lepton number (ELN) distribution $G_{\vec{v}} = \int _{-\infty}^{+\infty}d\omega\, g_{\omega, \vec{v}}$, and the equation for $\mathsf{S}_{\omega, \vec{v}}$ becomes essentially $\omega$-independent. For locally azimuth-symmetric ELNs, Eq.\eqref{eom1} becomes
\begin{align}\label{eom2}
\big(\partial_{t}+v\partial_{z}\big)\mathsf{S}_{v} = \mu_{0}\int_{-1}^{+1} dv{'}G_{v'}\left(1-vv{'}\right) \mathsf{S}_{{v}{'}} \times \mathsf{S}_{v}\,,
\end{align}
where $v$ is the radial velocity and $\mu_{0}$ is the collective potential. Initial conditions are ${\sf S}_{\omega,\vec{v}}\rvert^\text{ini}=+\hat{\sf e}_3$ and Eq.\eqref{eom2} is the same for all $\omega$, so $\nu$ and $\bar\nu$ have identical solutions. In our algebra, hereafter, $t=\mu_{0}t$ and $z=\mu_{0}z$, which are dimensionless. For concreteness, ELNs are taken to be piecewise-constant with one crossing at $v=0$,
\begin{align}\label{eq:gv}
G_{v}=\begin{cases}
    1, & \text{if $v>0$}\,,\\
    A-1, & \text{if $v<0$}\,,
  \end{cases}
\end{align}
and the lepton asymmetry $A=\int_{-1}^{+1} dv\,G_{v}$ takes values in $\{0.0, 0.2,\,0.4,\,0.5,\,0.6,\,0.9\}$.  For our numerical examples, we solve Eq.\eqref{eom2} with \mbox{$\mu_{0}=33\,\text{cm}^{-1}$}, corresponding to $n_\nu\approx5\times10^{33}\,\text{cm}^{-3}$. \bd{Periodic boundary conditions are assumed on \bd{$z\in \Delta r = (-1.5,+1.5)$\,cm}, treating this ``box'' as a part of a larger system. In lieu of ${\sf H}_\omega^\text{vac}$, the ${\sf S^{\perp}_{v}}[z,t=0]$ are explicitly seeded with amplitude $10^{-6}$ to start the flavor evolution. This choice plays a negligible role in deciding the final state; see the Supplemental Material (SM) for more details. The numerical methods are the same as in Ref.~\cite{Bhattacharyya:2020dhu}.}

\emph{Multipole~Diffusion.---} We define $\mathsf{M}_{n} = \int_{-1}^{+1}dv\,G_{v}L_{n} \mathsf{S}_{v}$ as the ${n}^\textrm{th}$ moment of $\mathsf{S}_{{v}}$, with $L_{n}[v]$ being the $n^\textrm{th}$ Legendre polynomial in $v$. In terms of $\mathsf{M}_{n}$, Eq.\eqref{eom2} becomes
\begin{align}\label{eom3}
\partial_{t}\mathsf{M}_{n}-\mathsf{M}_{0} \times \mathsf{M}_{n}= \partial_{z} \mathsf{T}_{n}-\mathsf{M}_{1} \times \mathsf{T}_{n}\,,
\end{align}
where $\mathsf{T}_{n} =  \frac{n+1}{2n+1}\mathsf{M}_{n+1}+\frac{n}{2n+1}\mathsf{M}_{n-1}$ that approximates to $\mathsf{M}_{n} + {\partial_{n} \mathsf{M}_{n}}/{(2n+1)}+{\partial^{2}_{n} \mathsf{M}_{n}}/{2}$ in the continuum limit of the discrete variable $n$~\cite{Raffelt:2007yz}.
After dotting Eq.\eqref{eom3} with $\mathsf{M}_{n}$ \bd{and averaging over $\Delta r$,  assuming it distributes over other operations}, one finds for large $n$:
\begin{equation}\label{eom4}
\partial_{t}\langle{M_{n}}\rangle  = \frac{\langle{M}_{1}\rangle}{2} \left(\partial^{2}_{n} \langle{M}_{n}\rangle  + \frac{1}{n}\partial_{n} \langle{M}_{n}\rangle \right)\,.
\end{equation}
\bd{The full derivation is given in the SM.} Here $\langle{M_{n}}\rangle$ denotes \bd{the spatially coarse-grained value} of $M_{n} = |{\mathsf{M}}_{n}|$. Eq.\eqref{eom4} is a diffusion-advection equation where $n$ plays the role of space and $\langle{M_{1}}\rangle$ of the diffusion constant.  $G_{v}$ and initial conditions for ${\sf S}_{v}$ are smooth in $v$, so that $\langle{M}_{n}\rangle$ are initially small for $n\gg1$. \bd{As time passes, the system \emph{diffuses} from low-$n$ to high-$n$ multipoles.} 

One can obtain an analytical solution to the above partial differential equation if $\langle{M}_{1}\rangle$ is approximately constant. First we note that Eq.\eqref{eom4} remains invariant under the scaling $n \rightarrow a n$ and $t \rightarrow a^{2}t$ with $a > 0$. Therefore, the solution for $\langle M_{n} \rangle$ can depend on $n$ and $t$ only through the scaling variable $\xi={n^{2}}/{{t}}$. Using $\xi$ as the independent variable, Eq.\eqref{eom4} becomes an ordinary differential equation, $2d^{2}_{\xi}\langle M_{n}\rangle +\big(1/\langle M_{1}\rangle + 2/\xi\big)d_{\xi}\langle M_{n}\rangle=0$. This has a solution $\langle M_{n}\rangle=c_{1}\,\text{Ei}\big[-{n^{2}}/\left({2\langle M_{1}\rangle t}\right)\big]+c_{2}$\,,
in terms of the exponential integral $\text{Ei}[x]=\int_{-\infty}^{x}dy\,e^{y}/y$. 
This solution, valid for large $n$, predicts {how} each $\langle M_{n}\rangle$, starting at $\langle M_{n}\rangle^{\text{ini}}$, grows exponentially, \bd{peaks at $t_n^\text{peak}\approx n^2/(2\langle M_{1}\rangle)$,} and asymptotes to $\langle M_{n}\rangle^{\text{fin}}$ at large times. The finite behavior at large $t$ is crucial to be able to truncate the multipole expansion. The solution shows that kinematic decoherence has a strong dependence on $\langle M_{1} \rangle$, which is initially $1-A/2$ for our ELNs. Thus, for small lepton asymmetry $A$ the effective diffusion coefficient $\langle M_{1} \rangle$ is larger. Further, shrinking of $\langle M_{1} \rangle$ results in less kinematic decoherence at later times, and as time progresses the system reaches an almost steady state with no further diffusion in multipole space. On the other hand for larger lepton asymmetry, i.e., smaller $\langle M_{1} \rangle^\text{ini}$, there is less diffusion and depolarization throughout.

\begin{figure}[!t]
	\centering
	\includegraphics[width=0.54\columnwidth]{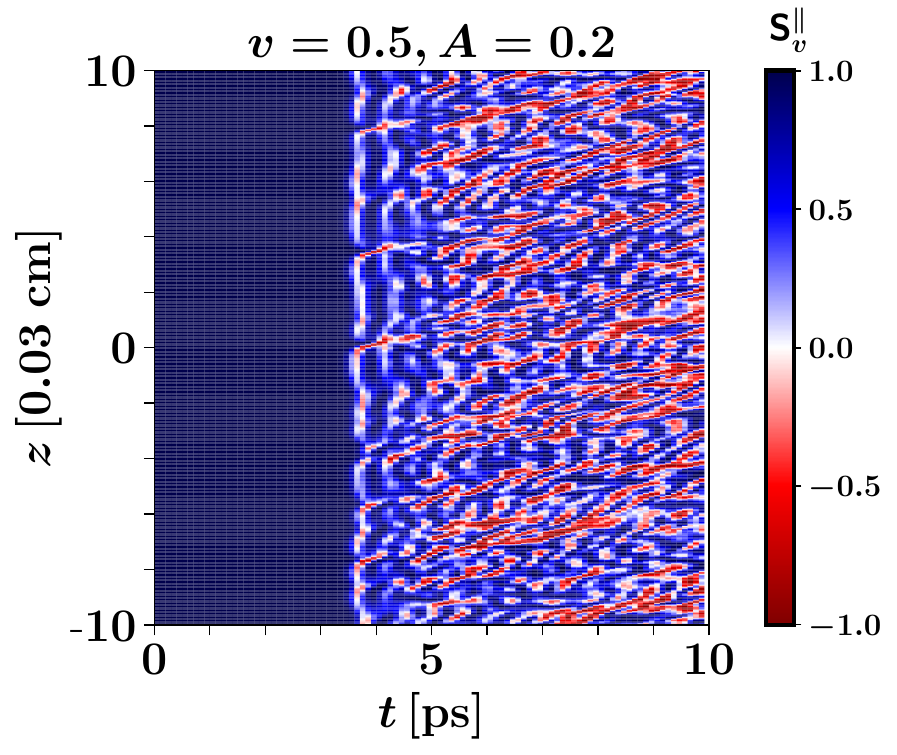}\hfill\includegraphics[width=0.46\columnwidth]{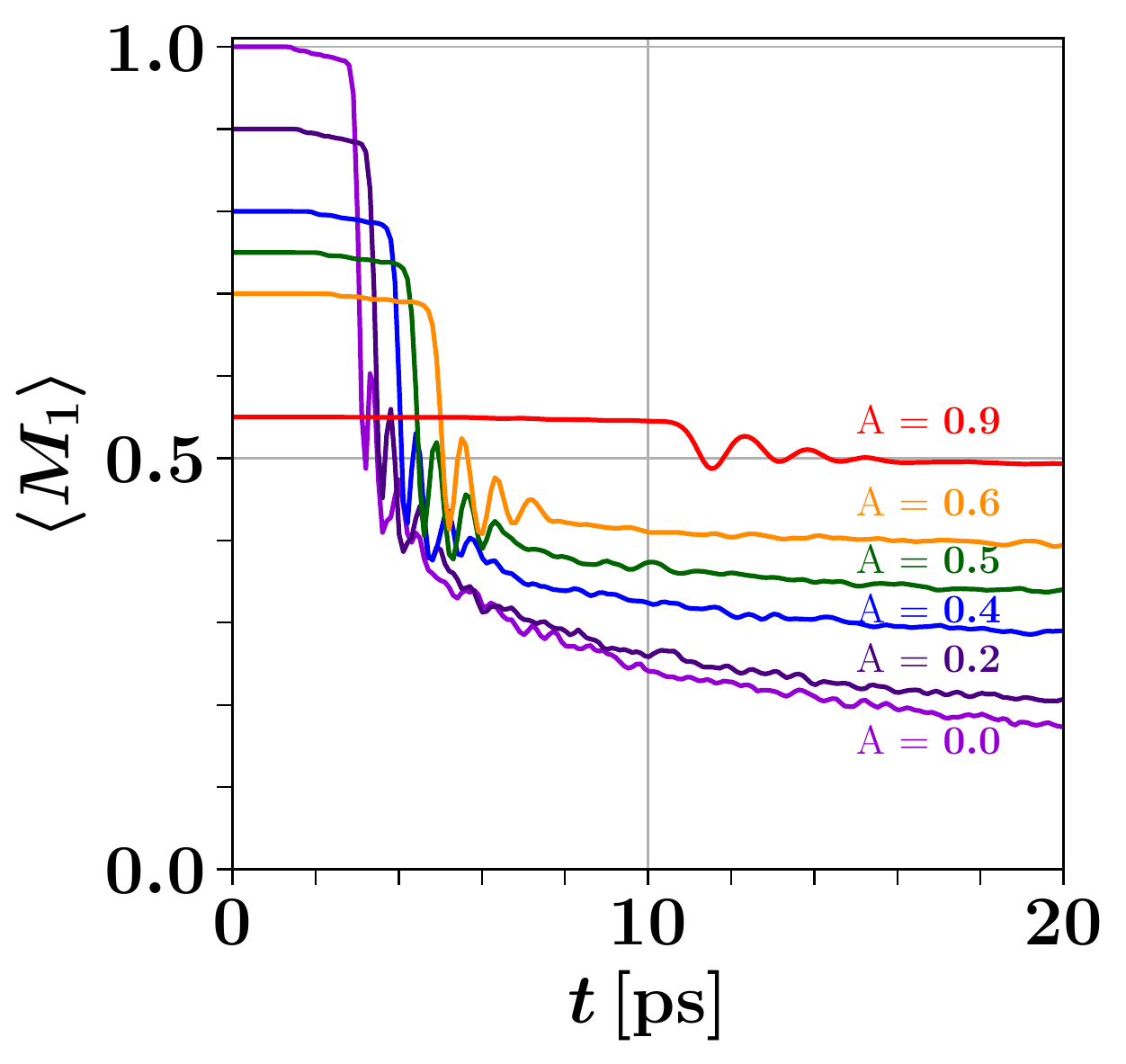}\\
     \includegraphics[width=0.95\columnwidth]{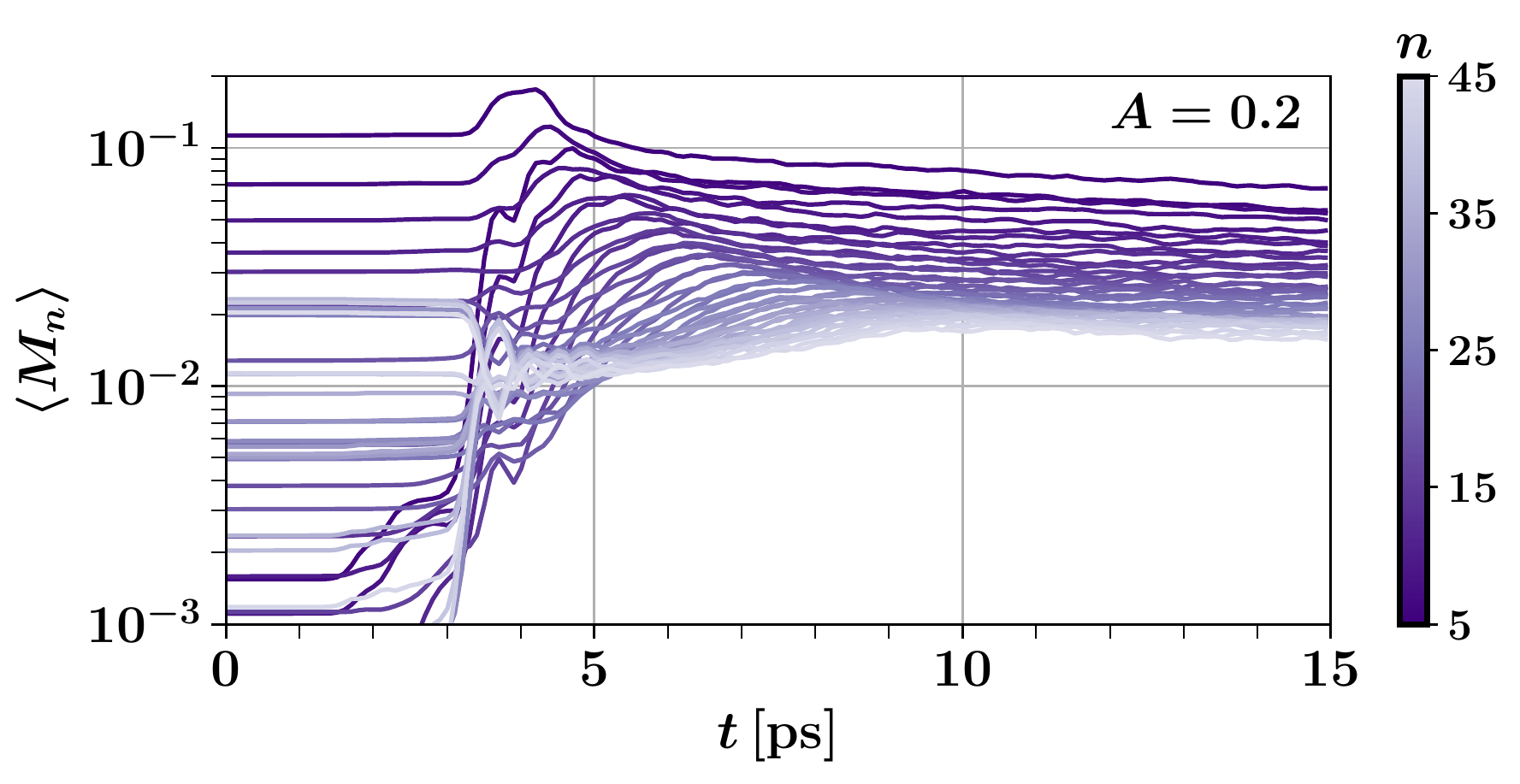}
	\caption{{\sc Multipole Diffusion:} Evolution of ${\sf S}^\myparallel_v$ for $v=0.5$ and $A=0.2$ (top left) and $\langle M_1\rangle$ for various ELNs (top right). Evolution of $\langle M_{n}\rangle$ for large $n$ and $A=0.2$ (bottom panel).}
	\label{fig3}
\end{figure}

\begin{figure}[!t]
	\includegraphics[width=0.51\columnwidth]{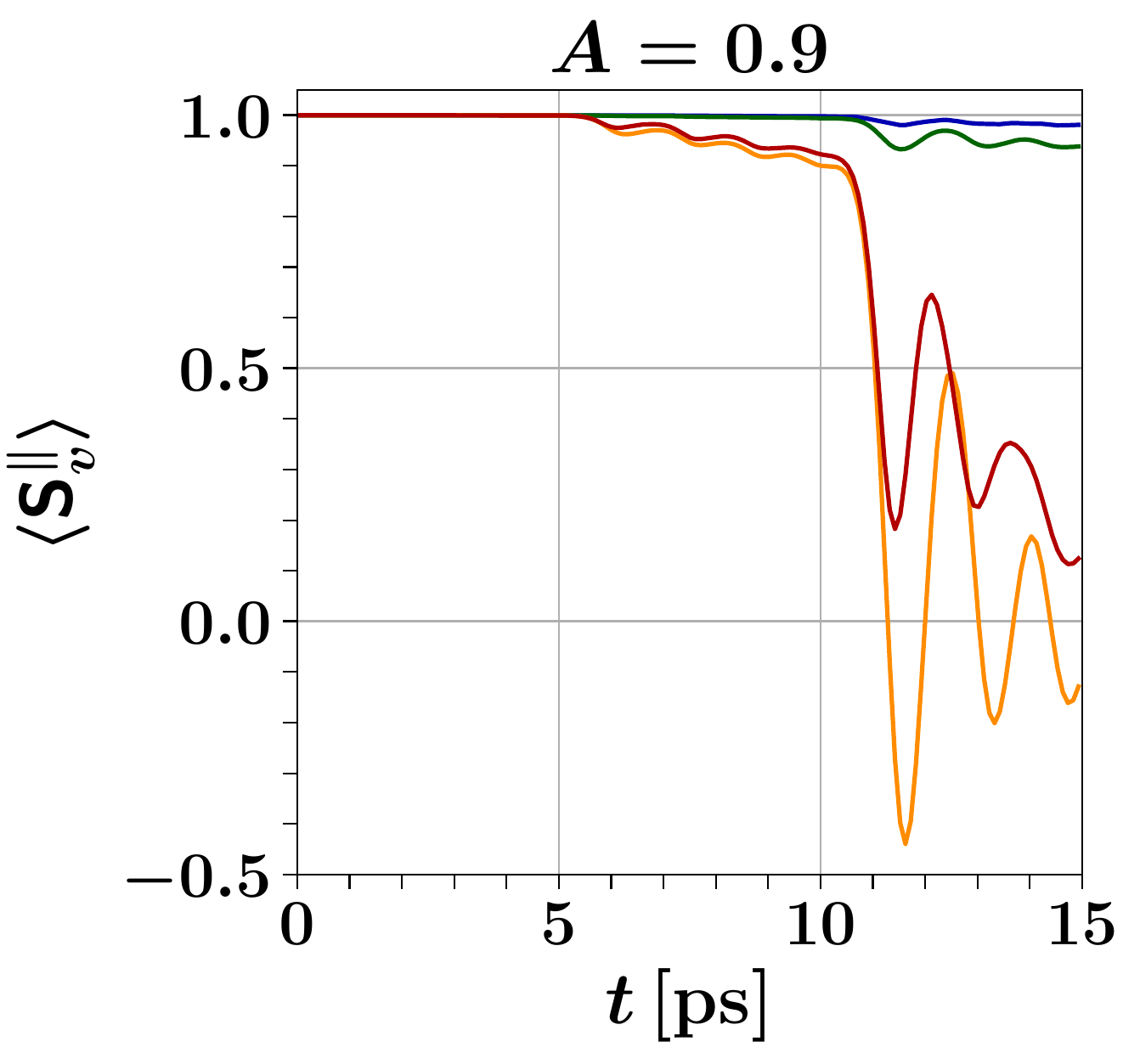}\,\includegraphics[width=0.467\columnwidth]{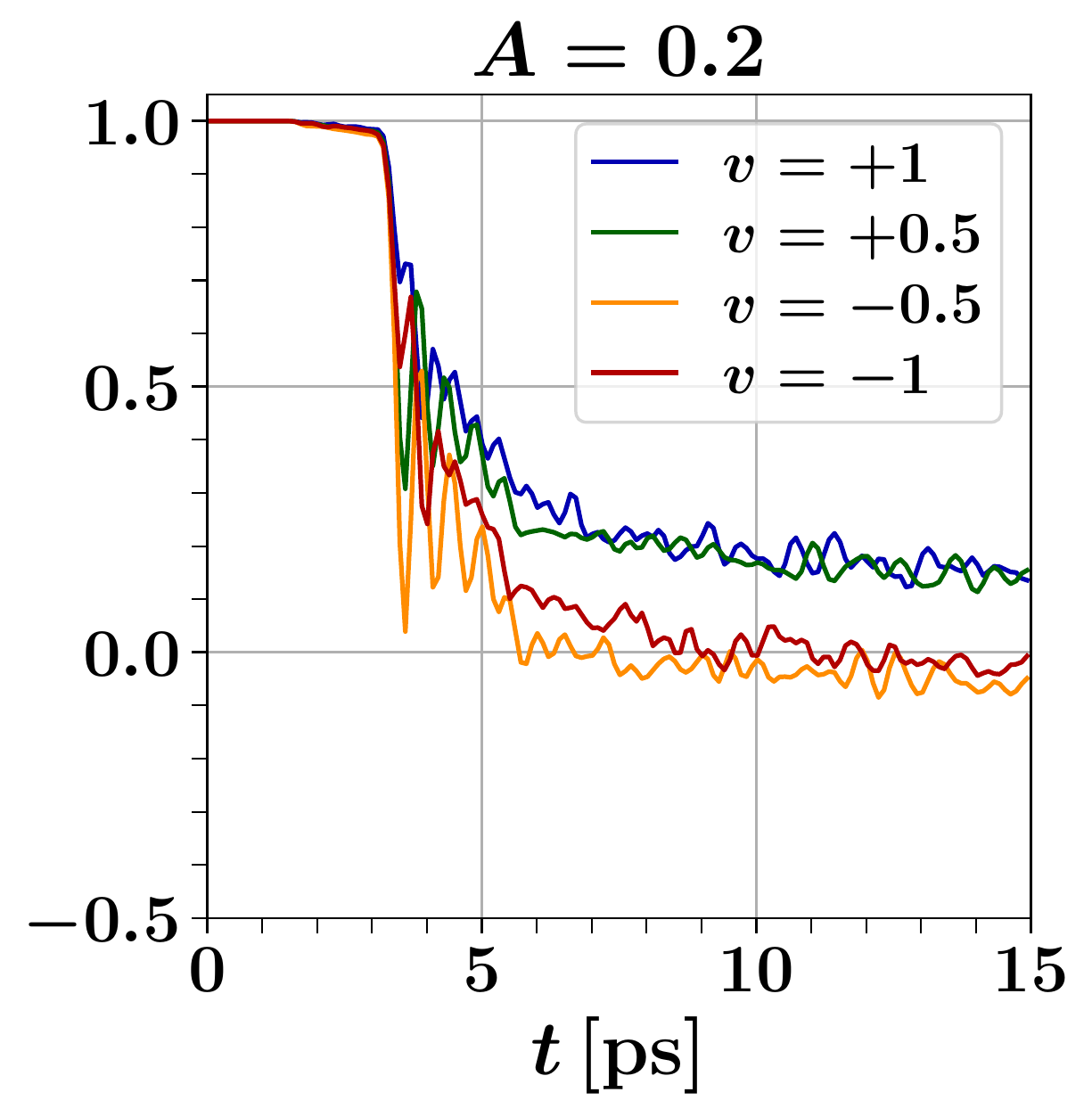}\\
	\hspace{0.5mm}\includegraphics[width=0.51\columnwidth]{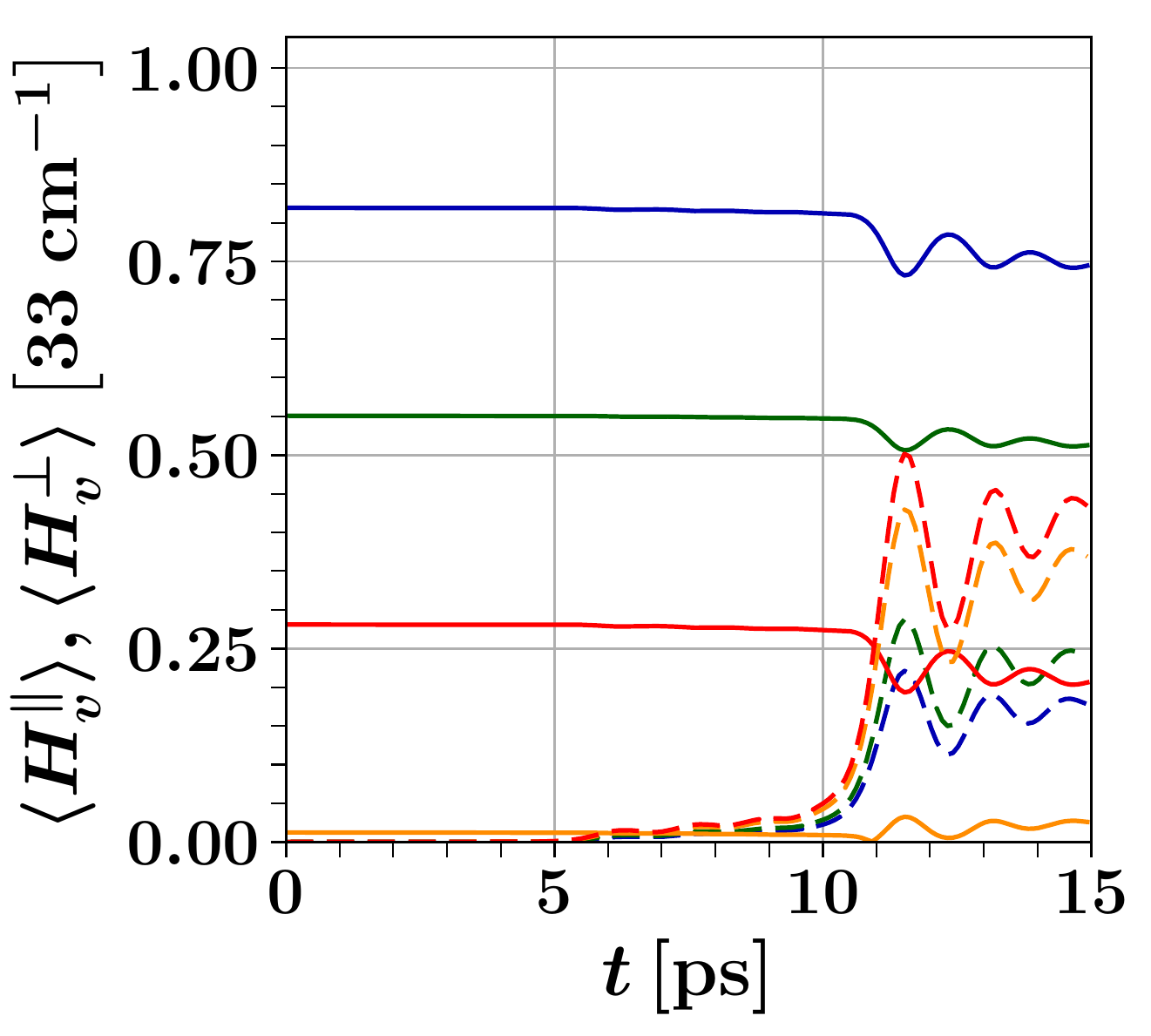}
	\includegraphics[width=0.467\columnwidth]{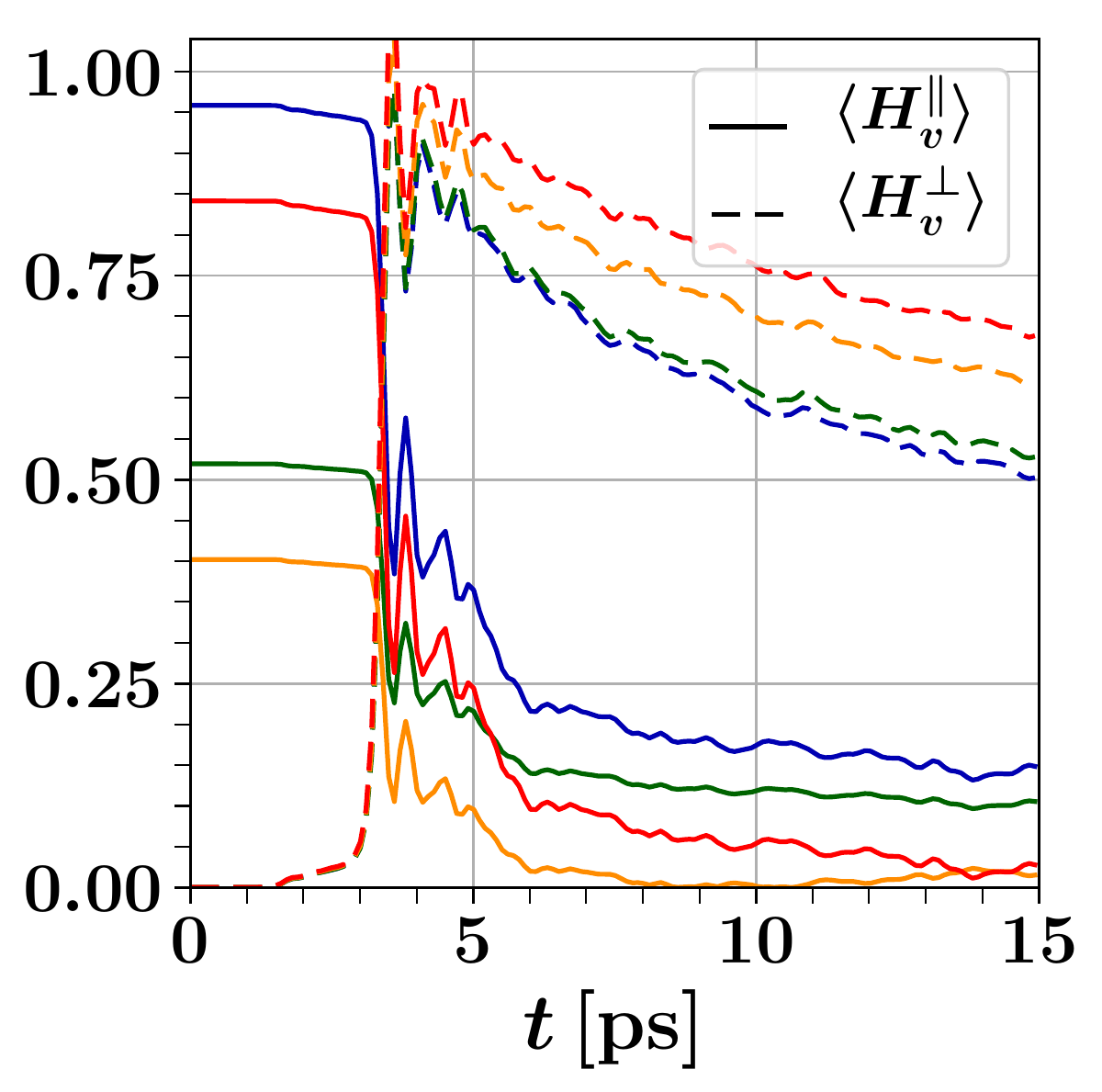}
    	\caption{{\sc Relaxation:} Evolution of $\langle {\sf S}^\myparallel_v\rangle$ for $v = \pm 1, \pm 0.5$ (top panels) for $A = 0.9$ (left) and $A = 0.2$ (right). $\langle{H}^{\myparallel}_v\rangle$ and $\langle{H}^{\perp}_v\rangle$, in solid and dashed lines, respectively (bottom panels).}
	\label{fig4}
\end{figure}

To verify the above analytical solution, we numerically solve Eq.\eqref{eom2} for our suite of ELNs. In Fig.\,\ref{fig3}, we show an illustrative result for ${\sf S}^{\myparallel}_{v}$, the $\langle M_{1} \rangle$ for all the ELNs, and various $\langle M_{n}\rangle$ for $A=0.2$. The top left panel shows how the flavor composition, even for a single $v$ mode, is scrambled within picoseconds and sub-mm distances. \bd{This time-scale depends logarithmically on the initial seed but the final state is insensitive to it.} In the right panel, we see $\langle M_{1} \rangle$ is approximately constant at early and late epochs, but decreases at $t\approx$\,$3.5$\,ps. We will explain the decrease in just a moment, but using the approximately constant $\langle M_{1} \rangle$ in our analytical solutions for $\langle M_{n}\rangle$, we find qualitative agreement with the numerical results shown in the bottom panel of Fig.\,\ref{fig3}. The sharp change in $\langle M_{1} \rangle^\text{ini}$ at $t\approx$\,$3.5$\,ps prevents a perfect agreement. Higher multipoles (fainter curves) rise, peak, and fall asymptotically, one-by-one, as predicted.


\emph{Transverse~Relaxation.---}
\bd{For the lower-$n$ multipoles, e.g., $\langle M_{0}\rangle$, $\langle M_{1}\rangle$, etc., the preceding discussion does not apply. Rather, comparing the top and bottom panels in Fig.\,\ref{fig4}, one sees that $\langle{\sf S}_v^{\myparallel}\rangle$ shrinks \emph{if} and \emph{when} $\langle{H}^{\perp}_v\rangle\approx\langle{H}^{\myparallel}_v\rangle$. We now explain this phenomenon. 
\emph{Naively}, the spatial average of Eq.\eqref{eom2} is $d_t \langle{\sf S}_v\rangle = \langle{\mathsf{H}}_v\rangle\times\langle{{\sf S}}_v\rangle$,
which can be visualized as a spin $\langle{{\sf S}}_v\rangle$ precessing around the magnetic field $\langle{\sf H}_v\rangle$. Note that $\mathsf{H}_{v}\approx-\left(\tfrac{1}{3}\mathsf{M}_{0}+v \mathsf{M}_{1}\right)$ in a frame corotating with the $\mathsf{M}_{0}$-$\mathsf{M}_{1}$ plane, for our choice of ELNs~\cite{Bhattacharyya:2020dhu}. Thus the length of $\langle{{\sf S}}_v\rangle$ ought to be constant. However, the length of $\langle{{\sf S}}_v\rangle$ in fact becomes smaller. Initially, $\langle{\sf S}_v\rangle$ is along $\hat{\sf e}_{3}$, and it starts tilting away due to the action of ${\sf H}^\text{vac}_{\omega}$. Considering its moments, $\langle\mathsf{M}_{0}\rangle$ is conserved, as $d_{t}\langle\mathsf{M}_{0}\rangle=0$, with $\langle\mathsf{M}_{0}^{\myparallel}\rangle=A$. On the other hand, $\langle\mathsf{M}_{1}\rangle$ has the motion of an inverted pendulum~\cite{Johns:2019izj,Bhattacharyya:2020dhu}. $\mathsf{M}_{1}$ tends to tip over, so that ${H}_{v}^{\myparallel}=|\tfrac{1}{3}A+v \mathsf{M}_{1}^{\myparallel}|$ becomes smaller as well. Eventually, when $\langle{H}^{\perp}_v\rangle \approx \langle{H}^{\myparallel}_v\rangle$, the $\langle{\sf S}_v\rangle$ makes a large precession angle and reaches the transverse plane. At this point, the averaging procedure does not factorize and ${\sf S}_v$ at different spatial locations relatively dephase and their coarse-grained transverse component $\langle{S}_v^\perp\rangle$ shrinks irreversibly~\cite{Bhattacharyya:2020dhu}. Thus the lengths $\langle{{S}}_v\rangle$ and $\langle M_{1}\rangle$ also become smaller. This mechanism of dephasing of transverse components is familiar as \emph{T2 relaxation} in the context of magnetic resonance imaging~\cite{doi:10.1002/jmri.23642}.}

Fig.\,\ref{fig4} shows that the $v<0$ modes, for which $\langle{H}^{\perp}_v\rangle$ overshoots $\langle{H}^{\myparallel}_v\rangle$, are depolarized completely, so \mbox{$\langle{\sf S}^{\myparallel}_{v<0}\rangle\to0$}. For $v>0$, the relaxation is less prominent, especially when $A$ is large. To zeroth order in $v$, one has $\langle{\sf S}^{\myparallel}_{v>0}\rangle\rvert^{\text{fin}}\approx A$, where we use $\langle{\sf S}^{\myparallel}_{v<0}\rangle\rvert^{\text{fin}}\to0$ and enforce conservation of lepton asymmetry. For our chosen form of $G_{v}$, with $A>0$ and a forward excess, it further implies that $\langle {\sf M}^{\myparallel}_{1}\rangle\rvert^{\text{fin}}\approx A/2$, as opposed to its initial value $1-A/2$. For ELNs with a backward excess and/or $A<0$, analogous arguments apply.

\emph{Depolarization.---}To quantify the effect of relaxation we define the depolarization factor as the relative reduction in the length of each Bloch vector,
\mbox{$f^\text{D}_{v}=\tfrac{1}{2}\big(1-{\langle{\sf S}_{v}^{\myparallel}\rangle^{\text{fin}}}/{\langle{\sf S}_{v}^{\myparallel}\rangle^{\text{ini}}}\big)$}. For flavor-pure initial conditions, ${\langle{\sf S}_{v}^{\myparallel}\rangle^{\text{ini}}}=1$.
As noted, $f^\text{D}_{v}$ is 0 (\textonehalf) when there is no (perfect) depolarization, and lies between \textonehalf~ and 1 if there is effective conversion to the other flavor. 

The extent of depolarization can be readily found. For positive lepton asymmetry, $A>0$, the negative velocity modes are almost completely depolarized, so clearly $f^\text{D}_{v<0}\approx$\,\textonehalf. For positive velocity modes the functional behavior of $f^\text{D}_{v>0}$ can be obtained by using the multipole expansion: $G_v {\sf S}^{\myparallel}_{v}\rvert^{\text{fin}}=\tfrac{1}{2}{\sf M}^{\myparallel}_{0}\rvert^{\text{fin}}+\tfrac{3}{2}\,v\,{\sf M}^{\myparallel}_{1}\rvert^{\text{fin}}+{\cal O}(v^{2})$, dropping the higher multipoles. As we found, $\langle {\sf M}^{\myparallel}_{0}\rangle=A$ is a constant in time but $\langle {\sf M}^{\myparallel}_{1}\rangle$ flips from $1-A/2$ to $A/2$. This brings us to the promised formula for the depolarization factor that was shown in Fig.\,\ref{fig2}:
 \begin{align}
 f^\textrm{D}_{v}  \approx 
 \begin{cases}
    \frac{1}{2}-\frac{A}{4}-\frac{3A}{8}\,v, & \text{if $v>0$}\,,\\
    \frac{1}{2}, & \text{if $v<0$}\,,
  \end{cases}
 \label{eq:fd}
 \end{align}
dropping the higher multipoles. For ELNs with a backward excess and/or $A<0$ the analogous formula for $f^\textrm{D}_{v}$ is easy to obtain using the mirror symmetries $+v\leftrightarrow-v$ and $+G^{(A<0)}_{v}\leftrightarrow-G^{(A>0)}_{-v}$ and a rescaling of $\mu_0$~\cite{Bhattacharyya:2020dhu}.

\emph{Summary {\textit{\&}} Outlook.---}
We have presented an analytical theory of fast neutrino flavor conversions in the nonlinear regime. We showed \emph{how}, as time passes, flavor differences over large ranges of velocity diffuse into variations over smaller velocity ranges, or equivalent ranges of emission angles, causing depolarization. Coarse-graining, by averaging over a small spatial volume and over small ranges of $v$, introduces loss of information that leads to an apparent arrow of time out of the time-reversible Eq.\eqref{eom1}. $f^\textrm{D}_{v}$ used in Eq.\eqref{eq:flux} must be understood in a spatially averaged sense. These features, including both $v>0$ and $v<0$ modes, are carefully verified using our state-of-the-art numerics~\cite{Bhattacharyya:2020dhu}. \bd{In contrast, without nonlinearity or coarse-graining no irreversible depolarization occurs and one finds wave solutions~\cite{Martin:2019gxb}.} We then showed that the epoch of T2 relaxation determines \emph{when} depolarization occurs, and the initial lepton asymmetry $A$ determines the rate of flavor depolarization. Finally, we gave a strategy and a formula for computing the \emph{extent of depolarization}, which is the ultimate outcome for fast collective oscillations pointed out by Sawyer~\mbox{\cite{Sawyer:2005jk, Sawyer:2008zs, Sawyer:2015dsa}}. 

\bd{Like the Landau-Zener formula~\mbox{\cite{Landau:1932,Majorana:1932ga,Stueckelberg:1932,Zener:1932ws}}, strictly applicable for a linearly varying density at a Mikheyev-Smirnov-Wolfenstein resonance~\cite{PhysRevD.17.2369, Mikheev:1987jp}, our depolarization formula gives a simple formula for the chosen class of ELNs. Both need further generalization in real-world applications. A difference is that depolarization, as long as it occurs, is irreversible. This final ``thermalized'' state is insensitive to microscopic details, e.g., the different purple lines in Fig.\,\ref{fig2} are for different initial seeds, depending only on conserved quantities like  $A$. See the SM. The key insight is to identify the role of coarse-graining and relaxation, which leads to this universal behavior.}

The neutrino flux after suffering fast conversions can be determined using the depolarization factor $f^\textrm{D}_{v}$. In a SN, these fluxes are responsible for heating and cooling processes~\cite{Colgate:1966ax}. The net heating rate $\dot{Q}$ that is responsible for shock revival depends on the product of cross section $\sigma_{\alpha}\propto E_{\alpha}^{2}$ and luminosity $L_{\alpha}\propto v\,E_{\alpha}F_{\alpha}$, with the $\bar{\nu}_{e}$ and ${\nu}_{e}$ dominating owing to their larger cross sections~\cite{Janka:2000bt}. It is clear that depolarization can change $\dot{Q}$, because $\bar{\nu}_{e}$ and ${\nu}_{e}$ energies move closer to that of $\nu_{x}$, and the increase proportional to $E_{\nu_x}^3/E_{\bar{\nu}_e}^3$ can be quite large~\cite{Dasgupta:2011jf}. Including the effects of subsequent slow collective oscillations~\cite{Duan:2006an, Dasgupta:2009mg}, MSW conversions, propagation and earth effects~\cite{Dighe:1999bi}, allows one to determine the final neutrino signal from a SN explosion. These can be measured at current and upcoming neutrino telescopes and may provide a remarkable way to directly test neutrino-neutrino interactions~\cite{GalloRosso:2017mdz,Seadrow:2018ftp,Capozzi:2018rzl}. These Standard Model interactions have never been directly tested in a laboratory. Of course, a variety of other particle physics and astrophysics information may be gleaned from such a signal~\cite{Horiuchi:2017sku,Simpson:2019xwo,Horiuchi:2017qlw,Yang:2018yvk,GalloRosso:2018ugl,Cherry:2019vkv,Ko:2020rjq,deGouvea:2020eqq}. In many such analyses, knowing $f^\textrm{D}_v$ is important. For the first time, our work provides this crucial input.

What lies ahead? For the more exciting, one could use this set-up as a test of possible secret neutrino-neutrino interactions~\cite{Dighe:2017sur}, that have been proposed as a solution to the Hubble tension~\cite{Cyr-Racine:2013jua,Archidiacono:2013dua}. Collective flavor conversions may also occur in the disk of merging neutron stars~\cite{Dasgupta:2008cu,Wu:2017drk,1815221}. These possibilities are not yet fully explored. Sticking to basics, however, several improvements, extensions, and applications are possible. Three-flavor effects were ignored here~\cite{Dasgupta:2007ws,Airen:2018nvp,Chakraborty:2019wxe,Capozzi:2020kge}. It will be interesting to see if our approach can be extended to include higher order terms in $v$ and $A$, break the azimuthal symmetry, and include more complicated ELNs.  These are important, but won't qualitatively change the picture we painted. As regards experiments, the diffuse SN neutrino background may soon become detectable~\cite{Beacom:2003nk,pmid30814722}, and hopefully the next galactic SN is not too far in the future~\cite{Abe:2016waf,Novoseltsev:2017gxc,Abi:2020lpk}. These effects may also have observable impact on the neutron star merger events at LIGO~\cite{GBM:2017lvd}. It is therefore of paramount importance that predictions for neutrinos are put on a firm footing and the experiments are well-prepared~\cite{Friedland:2018vry,Li:2020ujl}, so that we can reliably extract all the physics out of these once-in-a-lifetime events.

\pagebreak

\onecolumngrid

\acknowledgements

{\emph{Acknowledgements.---}
We thank A.\,Dighe, G. Raffelt, M.\,Rege, R.\,Sensarma for helpful comments on the manuscript, and K.\,Maji for pointing out T2 relaxation as a part of his electrodynamics assignment. We also thank members of the Collective Exchange Journal Club at MPI Munich and the CCAPP Journal Club for helpful comments. B.D. is supported by the Dept.\,of Atomic Energy (Govt.\,of~India) research project 12-R{\textit{\&}}D-TFR5.02-0200 and the Max-Planck-Gesellschaft through a Max Planck Partner Group.}

\appendix

\section*{Supplemental Material}

In this Supplemental Material we present a detailed derivation of Eq.5 and Eq.6 in the main text. We also provide more details of the numerical method and choice of the initial ``seed'' that kickstarts the flavor evolution.

\section{Derivation of the Coarse-grained Multipole Diffusion Equation}
In the main text we have defined $\mathsf{M}_{n} = \int_{-1}^{+1}dv\,G_{v}L_{n}[v]\,\mathsf{S}_{v}$ as the ${n}^\textrm{th}$ moment of the flavor Bloch vector $\mathsf{S}_{{v}}$, with $L_{n}[v]$ being the $n^\textrm{th}$ Legendre polynomial in $v$. Conversely, one has $\mathsf{S}_{v} = \sum_{n}\big(n+\frac{1}{2}\big)L_{n}[v]\,\mathsf{M}_{n}$. We remind that $v$ is the radial velocity and $\mu_{0}$ is the collective potential that has been set to unity so as to work in dimensionless time $t=\mu_{0}t$ and space $z=\mu_{0}z$. Multiplying Eq.\,(3) in the main text by $L_{n}[v]$ and integrating from -1 to +1 gives
\begin{align}
\partial_{t}\mathsf{M}_{n}-\mathsf{M}_{0} \times \mathsf{M}_{n}=\partial_{z} \bigg( \frac{n+1}{2n+1}\mathsf{M}_{n+1}+\frac{n}{2n+1}\mathsf{M}_{n-1}\bigg)-\mathsf{M}_{1} \times \bigg( \frac{n+1}{2n+1}\mathsf{M}_{n+1}+\frac{n}{2n+1}\mathsf{M}_{n-1}\bigg)\,,
\label{eq:eom2sm}
\end{align}
where two familiar identities obeyed by Legendre polynomials, $\int_{-1}^{+1}\,dv L_{n}[v]\,L_{m}[v]=\frac{2}{2n+1}\,\delta_{mn}$ and $\int_{-1}^{+1}\,dv\,v\,L_{n}[v]\,L_{m}[v] = \frac{2\left(n+1\right)}{\left(2n+1 \right)\left(2n+3\right)} \, \delta_{n+1, m}+\frac{2n}{\left(2n-1 \right)\left(2n+1\right)} \, \delta_{n-1, m}$, are used. 
Denoting the quantity in brackets in Eq.\,\eqref{eq:eom2sm} by $\mathsf{T}_{n}$, we get
\begin{align}
\partial_{t}\mathsf{M}_{n}- \mathsf{M}_{0} \times \mathsf{M}_{n}=\partial_{z} \mathsf{T}_{n}-\mathsf{M}_{1} \times \mathsf{T}_{n}\,.
\label{eq:eom3sm}
\end{align}
This is given as Eq.\,(5) in the main text.

We take a scalar product of the above equation with $\mathsf{M}_{n}$ to get
\begin{align}
\frac{1}{2}\partial_{t}{M}_{n}^{2}=\mathsf{M}_{n}\cdot\partial_{z} \mathsf{T}_{n}-\mathsf{M}_{n}\cdot(\mathsf{M}_{1} \times \mathsf{T}_{n})\,.
\label{eq:eom4sm}
\end{align}
Then we divide Eq.\,\eqref{eq:eom4sm} by $M_{n}$ to get
\begin{align}
\partial_{t} M_{n} -\mathsf{u}_n \cdot \partial_{z} \mathsf{T}_{n} = \left({\mathsf{M}_1 \times \mathsf{u}_n}\right)\cdot \mathsf{T}_{n}\,,
\label{eq:eom5sm}
\end{align}
where the unit vector $\mathsf{u}_n = {\mathsf{M}_n}/{M_n}$ points along $\mathsf{M}_n$. In deriving Eq.\,\eqref{eq:eom5sm} from Eq.\,\eqref{eq:eom4sm} we have used the vector identity $\mathsf{A} \cdot \left(\mathsf{B} \times \mathsf{C} \right) = -\left(\mathsf{B} \times \mathsf{A} \right)\cdot\mathsf{C}$. The discrete multipole variable $n$ can be considered as a continuous variable in the limit of large $n$. Using a Taylor-like expansion around $n$, with $\Delta n =1$, one has
\begin{align}
\mathsf{M}_{n\pm1}=\mathsf{M}_{n} \pm \Delta n\,\partial_{n} \mathsf{M}_{n}+\frac{1}{2}(\Delta n)^{2}\,\partial_{n}^{2}\mathsf{M}_{n}+\ldots\,.
\end{align}
Using the above approximation in the expression for ${\sf T}_n$ we get
\begin{align}
\mathsf{T}_{n}=\frac{n+1}{2n+1}\mathsf{M}_{n+1}+\frac{n}{2n+1}\mathsf{M}_{n-1}\approx\mathsf{M}_{n} + \frac{\partial_{n} \mathsf{M}_{n}}{2n}+\frac{\partial^{2}_{n} \mathsf{M}_{n}}{2}+\ldots\,,
\label{eq:largeN}
\end{align}
where we use $n\gg 1$, i.e., specifically $2n+1\approx 2n$ in the last step. Now, we use Eq.\,\eqref{eq:largeN} in Eq.\,\eqref{eq:eom5sm}, and coarse-grain by averaging over  $z\in(-L/2,+L/2)$, i.e., $\langle \cdots \rangle = \frac{1}{L}\int_{-L/2}^{+L/2} dz\,(\cdots)$, and approximate that the averaging procedure distributes in a phase-averaged sense over scalar and vector products, i.e., $|\langle {\sf A}\cdot{\sf B}\rangle| \sim |\langle {\sf A}\times{\sf B}\rangle| \sim \langle {A} \rangle \langle {B} \rangle$, as well as over derivatives with respect to $n$ and $t$, i.e., $\langle \partial_{n,t}{\sf A} \rangle=\partial_{n,t}\langle{\sf A} \rangle$, but not $z$; instead, the periodic boundary condition over $z$ gives $\langle \partial_{z}{\sf A}\rangle = {\sf A}|_{+L/2}-{\sf A}|_{-L/2}=0$, which removes the $z$-derivatives.
Altogether, this gives
\begin{align}
\partial_{t}\langle{M}_{n}\rangle=\frac{\langle{M}_{1}\rangle}{2} \bigg({\partial^{2}_{n} \langle{M}_{n}\rangle}+\frac{1}{n}\,{\partial_{n} \langle{M}_{n}\rangle}\bigg)\,.
\label{finalEqsm}
\end{align}
Eq.\,\eqref{finalEqsm} is the diffusion-advection equation given as Eq.\,(6) in the main text.

\section{Numerical Methods}
We developed our own code for solving Eq.\,(3) in the main text, previously documented in ref.\cite{Bhattacharyya:2020dhu}. Here we provide the same information for easy reference. In our code, we discretize both the spatial direction as well as the velocities, considering $N_{z}$ number of spatial and $N_{v}$ number of velocity bins, to get a total of $3\,N_{z}\,N_{v} $ coupled nonlinear ODEs, where the factor of 3 comes due to the 3 elements of each polarization vector. It solves these coupled nonlinear ODEs as a function of time using {\tt python}'s {\tt zvode} solver, which is a complex-valued variable-coefficient ordinary differential equation solver in {\tt python} which implements Backward Differentiation Formula for doing numerical integration. The spatial derivatives at each spatial point are computed using {\tt python}'s {\tt scipy.fftpack.diff} package, which uses the Fast Fourier Transform method for calculating derivatives.

For the computations shown in this paper, we choose a periodic boundary condition in space with a period equal to the box size $L$, such that, for any time $t$ and any velocity mode $v$, our solution satisfies
\begin{equation}\label{40}
\mathsf{S}_{v} [z,t] = \mathsf{S}_{v}[z+L, t] \,.
\end{equation}
We choose initial conditions such that all the neutrinos, with any velocity, are emitted as purely electron flavored states at every point in space. In the absence of ${\sf H}^\text{vac}$,  the flavor evolution is triggered by an external perturbation, described in Eq.\eqref{16}-\eqref{18}, with amplitude $\epsilon = 10^{-6}$ that is used as an initial seed for the transverse components of the polarization vectors, for every velocity mode and at all spatial locations.

Specifically, we choose a 1D box of size $\mu_0L = 115$  and discretize it into $N_z = 2^{12}$ spatial bins to solve Eq.\,(3) in the main text numerically up to a time $\mu_0 t = 50$. As mentioned in text, we use $\mu_0=33\,\text{cm}^{-1}$. We discretize the ELN distributions into $N_v = 2^{7}$ velocity modes, giving rise to a total of $3\times 2^{12} \times 2^{7} = 1572864$ coupled ODEs. These choices are optimized to obtain an accuracy of about $10^{-4}$ and precision of about $10^{-8}$, as shown previously in Fig.\,{9} of ref.\cite{Bhattacharyya:2020dhu}. Fig.\,3 in the main text clearly indicates that the system becomes nonlinear already at $t \approx 3.5$\,ps. To ensure that we are deeply in the nonlinear regime so that the system becomes completely thermalized and maintains a steady state, we run our code until $t=50$\,ps and show our final-state results in Fig.\,2 in the main text at $t=50$\,ps. 

\subsection{Role of the Initial Seed}
For the numerical solutions in the main text we have dropped the mass-mixing Hamiltonian ${\sf H}_\text{vac}$ that starts the evolution, and chosen to kick-start the flavor evolution with an explicit seed for ${\sf S}_{v}^{\perp}$. The late-time solution is insensitive to the choice of seed, as long as the fast instability occurs and reaches nonlinearity. Here we provide a pedagogical discussion of this issue.

Triggering the instability with ${\sf H}_\text{vac}$ would work just fine with only a small change: the initial seed would be approximately homogeneous in space and thus the instability is dominated by the corresponding $k=0$ mode. This can be slow and numerically expensive. Also, it does not capture the physical possibility that there are spatial fluctuations that would trigger other $k$-modes.

\begin{figure*}[!t]
	\includegraphics[width=0.3\textwidth]{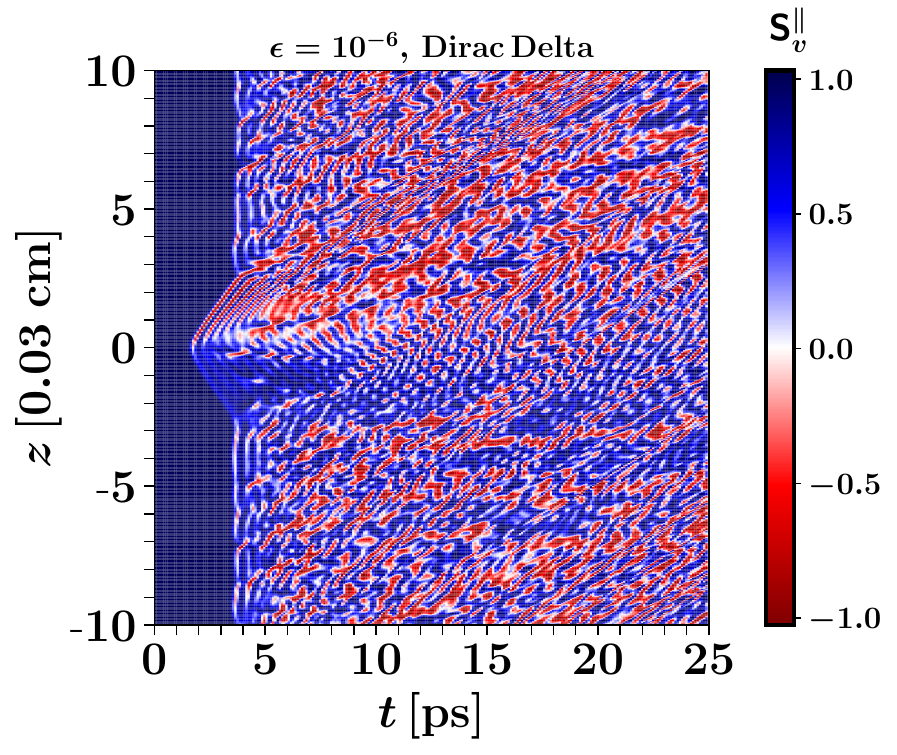}\,\includegraphics[width=0.3\textwidth]{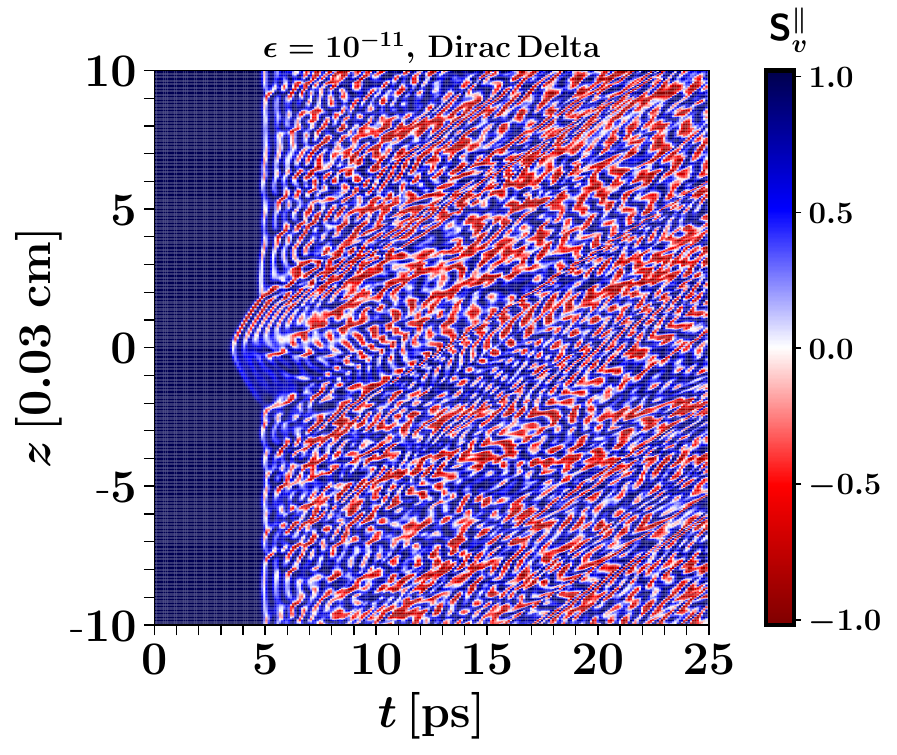}\,
	\includegraphics[width=0.3\textwidth]{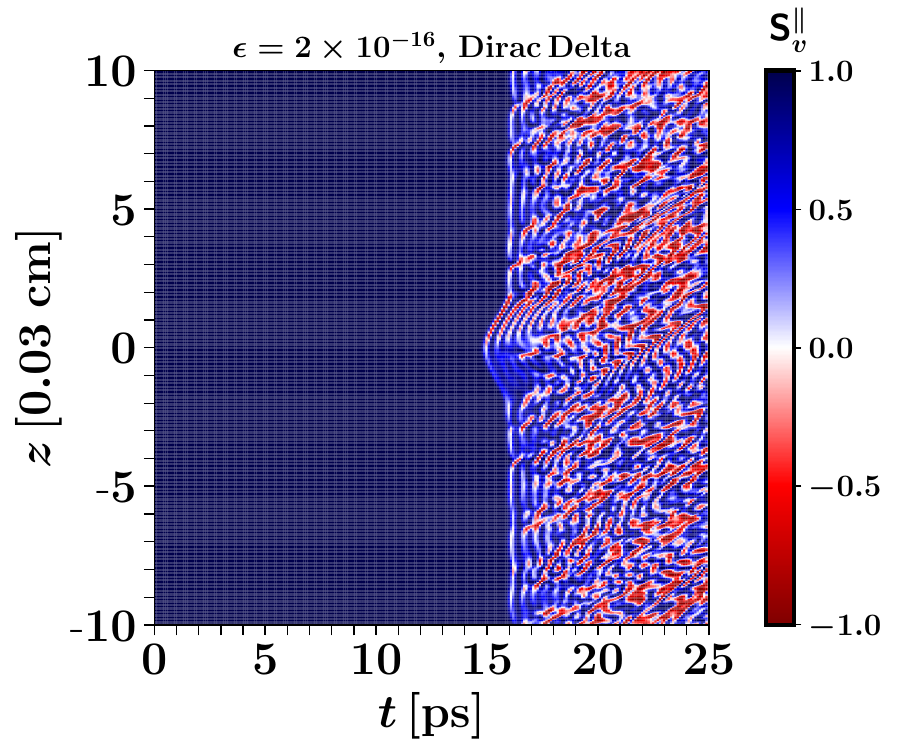}
    	\caption{{\sc Dirac Delta Seed:} As we reduce $\epsilon$, i.e, the amplitude  of the seed at $z=0$, the overall evolution is logarithmically delayed. The mode shown is $v=0.5$ for $A=0.2$, and the three panels correspond to seeds of amplitude $10^{-6}$, $10^{-11}$, and $2\times10^{-16}$ at $z=0$. An edge-like feature appears at 3.5\,ps, 5\,ps, and 16\,ps, respectively, due to numerical creation of pseudo-seeds at $z\neq0$ even if none were assigned originally. Given the unstable equations, these autonomously generated pseudo-seeds also grow and reach nonlinearity. This is clear from the last panel, where the true seed is close to the numerical precision ($10^{-16}$), so that the light-cone feature becomes less prominent and all points in $z$ take almost the same time to reach nonlinearity. Note also that the pseudo-seeds are proportional to the true seed because ${\sf S}^{(1,2)}[z\neq0,t\to0^{+}]\propto {\sf S}^{(3)}[z\neq0,t\to0]\,{\sf S}^{(2,1)}[z=0,t\to0]\propto\epsilon$.}
\label{fig1sm}
\end{figure*}

As for the choice of initial seeds, there are infinitely many choices. A seed that is constant over $z$ is exactly like starting the evolution with ${\sf H}_\text{vac}$ and offers no advantage. Putting a seed at $z=0$ only and nowhere else, i.e. ${\sf S}_{v}^{(1,2)}[z,t=0]\sim\epsilon\,\delta(z)$, is equivalent to computing the Green's function for the problem. It is a well-motivated choice. The results for this choice, $\lvert{\sf S}_{v}^{(1)}[z,t=0]\rvert=\lvert{\sf S}_{v}^{(2)}[z,t=0]\rvert=\epsilon\,\delta(z)$, are shown in Fig.\,\ref{fig1sm} for $v=0.5$ and $A=0.2$. The light-cone-like feature arises because there is a predominant seed at $z=0$. However, it may be confusing why there is a flat edge-like feature extending outside the light-cone. Immediately after $t=0$, small pseudo-seeds are generated at all $z\neq0$  due to numerical effects, e.g., finite numerical precision. These are extremely small errors and would be normally harmless. However, our chosen equations have an instability and these pseudo-seeds grow just like the true seed. So, the solution grows even in spatial locations where it was not supposed to. We will see that the long-term solution that we are interested in remains unaffected by this. Nevertheless this may be unpalatable.

\begin{figure*}[!bthp]
\centering
\includegraphics[width=0.3\textwidth]{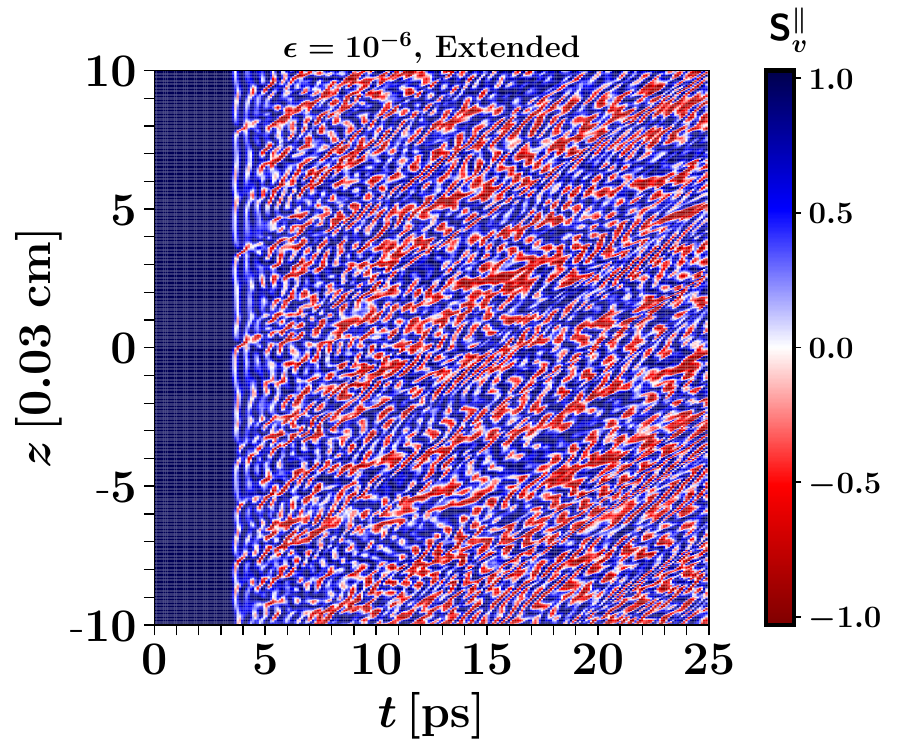}\,
    	\caption{{\sc Extended Seed:} If there are seeds at every $z$ there is no visible light-cone. The seed for this figure, which is identical to top-left panel of Fig.\,3 in the main text, is given by Eqs.\,\eqref{16}-\eqref{18}. The mode shown is $v=0.5$ for $A=0.2$, and the amplitude of the seed is $\epsilon=10^{-6}$.}
\label{fig2sm}
\end{figure*}

We therefore find it preferable to use initial conditions where there are seeds all over the periodic box. Now one only gets minuscule corrections to the pre-existing $\epsilon$-seed everywhere, that introduces no new instabilities and does not change the solution at a measurable level. However, setting the \emph{same} seed for all $z$ would select the $k=0$ mode only. Instead, to trigger a wider array of $k$-modes, one can consider the following initial condition:
\begin{align}\label{16}
{\sf S}_{v}^{(1)}[z,t=0]=\epsilon\cos\varphi[z]\,, 
\end{align}
and 
\begin{align}\label{17}
{\sf S}_{v}^{(2)}[z,t=0]=\epsilon\sin\varphi[z]\,,
\end{align}
where $\varphi[z]$ is chosen as any function of $z$ that obeys the periodic boundary condition. We take 
\begin{align}\label{18}
\varphi[z]=\frac{1}{N_{z}}\sum_{j=0}^{N_{z}-1}\cos\bigg[\frac{2\pi\,{j}\,z}{L}\bigg]\,,
\end{align}  
so that akin to the Dirac delta seed we seed all allowed modes in the range $k=0$ to $2\pi(N_z-1)/L$. However, now one has the dual advantage of triggering a wide range of $k$-modes and not being subject to large relative numerical errors at any spatial point. The results for this, again for $v=0.5$ and $A=0.2$, are shown in Fig.\,\ref{fig2sm}. No light-cone is seen with this choice because all spatial locations have a seed and grow roughly similarly.

\begin{figure*}[!bthp]
	\includegraphics[width=0.4\textwidth]{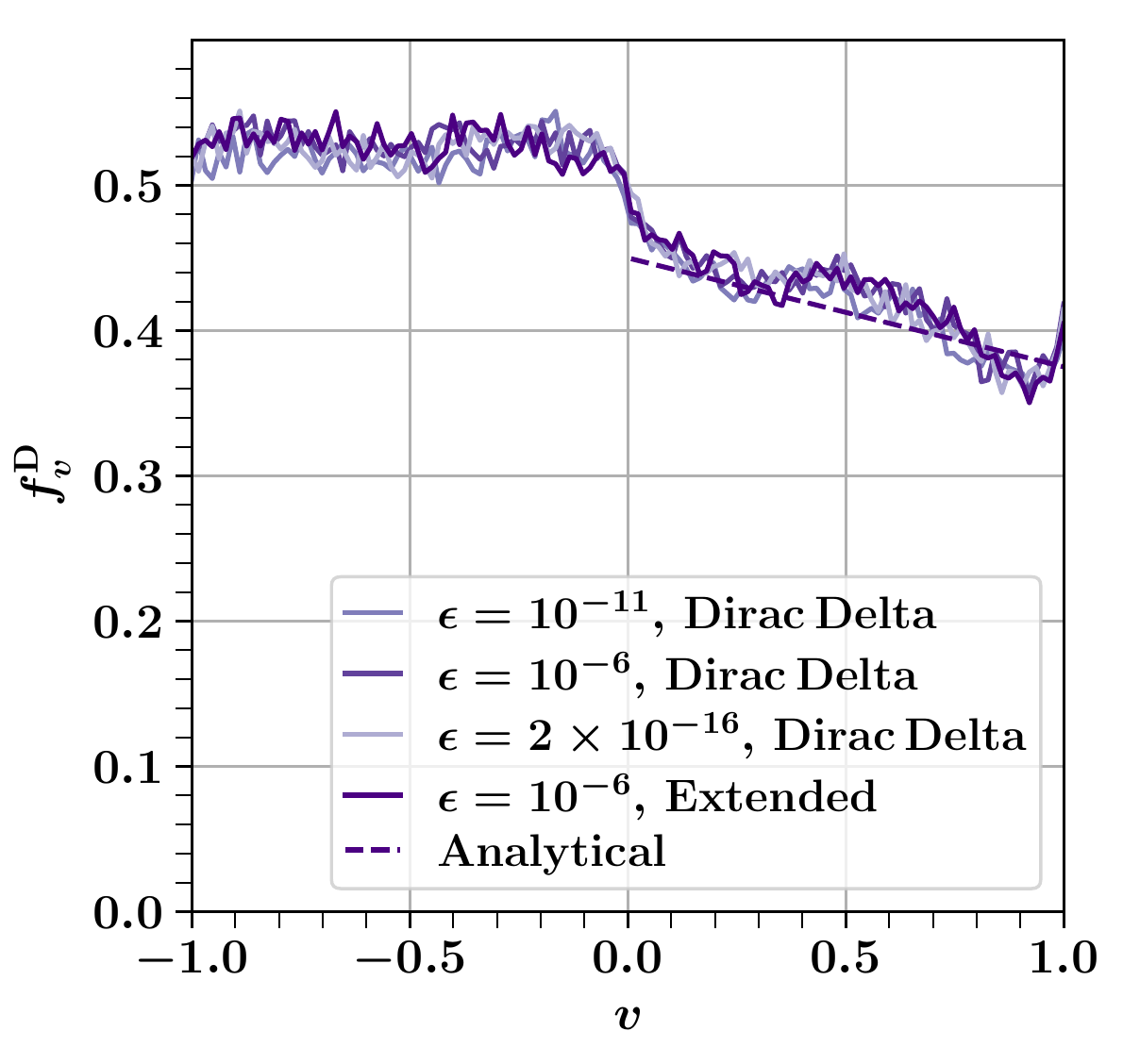}
    	\caption{{\sc Universality of Depolarization:} Analytical (dashed) and numerical (solid) results for coarse-grained $f^\textrm{D}_v$, as a function of the radial velocity, $v=\cos\theta$, same as Fig.\,2 of the main text. The different lines correspond to the different choice of initial seeds shown above. As evident, depolarization is universal and does not appreciably depend on the nature of seeds.}
	\label{fig3sm}
\end{figure*}

To demonstrate that there is no dependence of the depolarization factor on the seed, in Fig.\,\ref{fig3sm} we show the depolarization factor at $t=50$\,ps for the $A=0.2$ case, with a variety of initial seeding choices given above. As one can see, the coarse-grained analytical result, Eq.\,(7) of the main text, is an equally good fit to the numerically extracted $f_{v}^\text{D}$ for all the cases. The physics of this is obvious: the seeds have a logarithmic influence on \emph{when} the flavor evolution starts and becomes nonlinear, and \emph{which} $k$-modes are triggered, etc., but the long-term coarse-grained solution  does \emph{not} remain sensitive to these details. It is no different from why the macroscopic thermodynamic variables such as pressure or temperature have no relation to the microscopic initial conditions for the gas molecules.   

\begin{figure*}[!h]
	\includegraphics[width=0.5\textwidth]{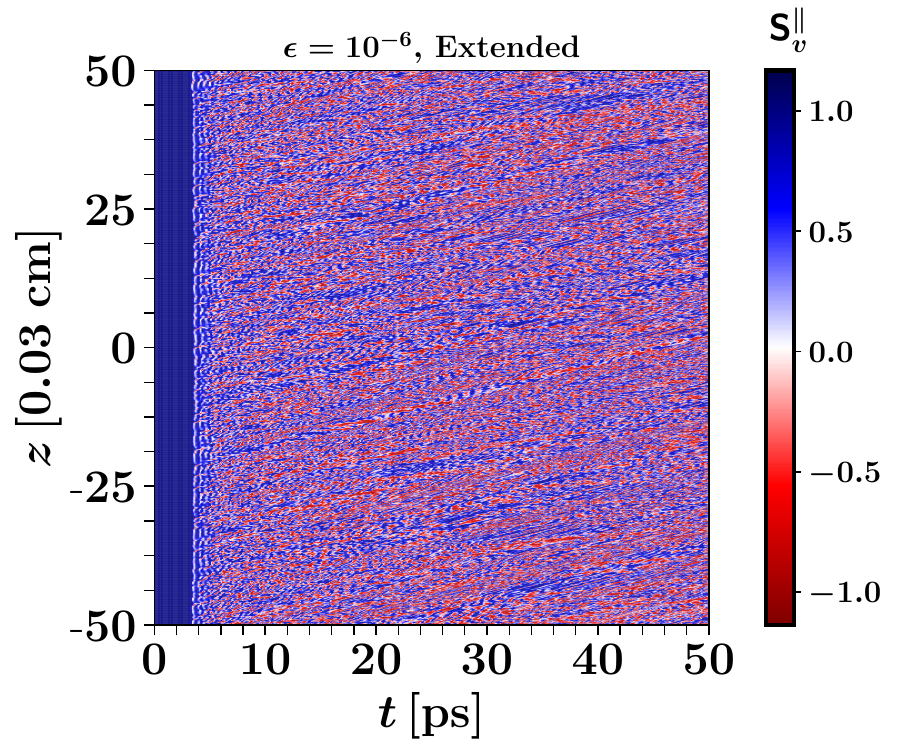}
    	\caption{{\sc Complete Flavor Evolution:} The figure is identical to top-left panel of Fig.\,3 in the main text, except for the larger range of time and space that is shown here. Substantial evolution occurs at $\approx3.5$\,ps and depolarization starts around $5$\,ps. After $t\sim10$\,ps the coarse-grained flavor composition is approximately steady. The diagonal streaks that one can see are because the velocity mode $v=0.5$ is shown here. For other modes one sees similar streaks but at a different slope corresponding to their $v$.}
	\label{fig4sm}
\end{figure*}

For completeness, in Fig.\,\ref{fig4sm} we also exhibit the complete flavor evolution, until $t=50$\,ps in the entire box, for the $v=0.5$ and $A=0.2$ case. The top-left panel of Fig.3 in the main text is a zoom-in of this plot to show the features more clearly. 

\bibliographystyle{JHEP}
\bibliography{references}

\end{document}